# PHYSICAL UNCLONEABLE FUNCTION HARDWARE KEYS UTILIZING KIRCHHOFF-LAW-JOHNSON-NOISE SECURE KEY EXCHANGE AND NOISE-BASED LOGIC

LASZLO B. KISH [(1)], CHIMAN KWAN [(2)]

[(1)] *Texas A&M University, Department of Electrical and Computer Engineering, College Station, TX 77843-3128, USA; email: Laszlo.Kish@ece.tamu.edu ; sunil@ece.tamu.edu*

[(2)] *Signal Processing Inc, 13619 Valley Oak Circle, Rockville, MD 20850, USA*

**Abstract.** Weak uncloneable function (PUF) encryption key means that the manufacturer of the hardware can clone the key but anybody else is unable to so that. Strong uncloneable function (PUF) encryption key means that even the manufacturer of the hardware is unable to clone the key. In this paper, first we introduce an "ultra"-strong PUF with intrinsic dynamical randomness, which is not only not cloneable but it also gets renewed to an independent key (with fresh randomness) during each use via the unconditionally secure key exchange. The solution utilizes the Kirchhoff-law-Johnson-noise (KLJN) method for dynamical key renewal and a one-time-pad secure key for the challenge/response process. The secure key is stored in a flash memory on the chip to provide tamper-resistance and non-volatile storage with zero power requirements in standby mode. Simplified PUF keys are shown: a strong PUF utilizing KLJN protocol during the first run and noise-based logic (NBL) hyperspace vector string verification method for the challenge/response during the rest of its life or until it is re-initialized. Finally, the simplest PUF utilizes NBL without KLJN thus it can be cloned by the manufacturer but not by anybody else.

## 1. Introduction

Intellectual Property (IP) protection of hardware designs such as FPGA and IC has been becoming important in recent years. It has been estimated that US$100 billion of global IT industry revenue has been lost due to counterfeiting [1]. Attackers can easily tap the bit stream containing hardware configuration information and hence clone the hardware system.

To protect IP of hardware systems, researchers developed a physical unclonable function (PUF) that is embodied in a physical structure and is easy to evaluate but hard to predict even given the exact manufacturing process that produced it [2]. Earliest research of PUF started in 1983 [3] and 1984 [4,5]. The term PUF (physical unclonable function) was coined in 2001 [6] and 2002 [7]. In [7], the first integrated PUF was produced in the same electrical circuit.

In PUF, a challenge $C_i$ and a response $R_i$, ($i = 1,2, ... , Q$) called a challenge-response pair (CRP) is needed. That is, a PUF is considered as a function that maps challenges to responses [1,2]. For PUF to work, there are 3 conditions [1]:

(*i*) It is required that a response $R_i$ (to a challenge $C_i$) gives only a negligible amount of information on another response $R_j$ (to a different challenge $C_j$) with $i \neq j$ ;

(*ii*) Without having the corresponding PUF at hand, it is impossible to come up with the response $R_i$ corresponding to a challenge $C_i$, except with an exponentially low probability (versus the bit-number characterizing the PUF); and

(*iii*) *It is assumed that PUFs are tamper evident*.

We note in passing that there is a frequent misunderstanding of the meaning of PUF. Particularly, inspired by PUF systems utilizing random optical media, *it is often believed that the PUF is defined by the technical difficulties of accessing and characterizing the physical structure of a PUF device*. That is not a PUF property but *only tamper-proof characteristics*, which is already assumed in condition (*iii*).

Thus, the technical difficulties of cloning a PUF by physical means belongs to condition (*iii*), which is guaranteed per-definition, and the no-cloning condition is strictly defined by conditions (*ii*) and (*iii*). This does not mean that tamper-evidence is not a serious problem in PUF development however PUF means a mathematical robustness against extracting the challenge-response set by observing the allowed number of challenges and the corresponding responses.

Concerning the randomness they utilize, there are 2 types of PUFs. The first category uses explicitly-introduced randomness. Examples include optical PUF and coating PUF [2]. One key advantage of this type of PUF is that a much greater ability to distinguish devices from one another and have minimal environmental variations compared to PUFs that utilize intrinsic randomness. The second type is the PUF using intrinsic randomness. Typical examples include delay PUF, SRAM PUF, Butterfly PUF, Bistable Ring PUF, and Magnetic PUF. Advantages of PUF include the following. First, there is no need for a dedicated power-source (e.g., a battery) to keep the key memory alive. Second, data remanence issues are eliminated. Third, additional physical security measures to protect the memory device are minimized.

However, it has been pointed out in [8] that all of the existing PUF in the literature still lack statistical uniqueness and robustness with respect to environmental conditions. In addition, the ability of the PUF to reliably and repeatedly generate a unique PUF of a minimum 256 bits in length still needs significant research. Finally, a paper [9] points out that many PUFs, including standard Arbiter PUFs and Ring Oscillator PUFs of arbitrary sizes, and XOR Arbiter PUFs, Lightweight Secure PUFs, and Feed-Forward Arbiter PUFs of up to a given size and complexity, are all vulnerable to numerical modeling attacks.

PUF can be classified also according to its strength. *Weak* PUF key can be cloned by its maker but by anybody else. *Strong* PUF cannot be cloned even by its maker. However, even stronger PUF, ultra-strong ones can be realized, which we will show below.

In this paper, we will introduce an *ultra-strong*, a *strong* and a *weak* PUF hardware key. These systems use two tools of noise-based informatics: the Kirchhoff-law-Johnson-noise (KLJN) unconditionally secure key exchange method [9-14], and noise-based logic (NBL) [16-26].

In Section 2, we briefly describe the KLJN system [9-14] and the required version of NBL tool [26].

In Section 3, we introduce a novel PUF key with intrinsic randomness. It maybe called "*ultra-strong*" PUF because not only it cannot be copied by its maker but during each use it is randomly changing by generating and sharing a new independent unconditional secret with the lock thus, if a counterfeiter accidentally succeeds to make a clone (this has exponentially low probability), that clone will be useless after next time the legal user applies the PUF key. The generation and regeneration of the intrinsic randomness in the key utilizes the unconditionally secure key distribution based on the Kirchhoff-law-Johnson-Noise (KLJN) method [9-14], which offers comparable or higher security than quantum key exchange (QKD) schemes do [13]. To generate the challenge/response a one-time-pad is used, which gets renewed during each use of the PUF key.

In Section 4, we present a simplified version of the PUF key described above, which is still "strong", that is even the manufacturer is unable to clone it after its first use. The key refreshment is used only first time when the PUF key is applied to the lock. Then during the subsequent applications the secure key does not act as a one-time-pad anymore. The challenge-response system utilizes noise-based logic (NBL) [16-25] hyperspace vectors [19] and the string verification method [26] based on that.

In Section 5, we present the simplest PUF utilizing noise-based informatics: a key that can be cloned by the manufacturer (this means it is not a strong version) however it cannot be cloned by anybody else. This system contains a secret key installed by the manufacturer. The challenge-response system utilizes the hyperspace vector [19] of noise-based logic (NBL) [16-25] and the string verification method [26] based on that.

## 2. Brief description of the tools: the KLJN secure key exchange and noise-based logic

*2.1 The KLJN key exchange method with information theoretic (unconditional) security*

The Kirchhoff-law-Johnson-noise (KLJN) secure key exchange scheme [9-14] was proposed in 2005 [9,10]. It is a statistical/physical competitor to quantum key exchange and its security is based on Kirchhoff's Loop Law and the Fluctuation-Dissipation Theorem. More generally, it is founded on the Second Law of Thermodynamics, which indicates that the security of the ideal scheme is similarly strong as the impossibility to build a perpetual motion machine (of the second kind).

The basic KLJN system [9,13] is shown in Figure 1. Two identical pairs of resistors ($R_L$ and $R_H$) and two binary switches with corresponding states (representing bit values) are used, one at Alice's side and one at Bob's. Depending on the state of a switch, which is randomly selected at the beginning of each clock cycle, one of resistors (with its own Johnson noise generator) is connected to the cable. Thus there are four possible states in which the two binary switches can be in: *LL*, *LH*, *HL*, and *HH*. The two mixed states, *LH* and *HL*, have identical mean-square channel noise voltage $U_{ch}$ and current $I_{ch}$ amplitudes, which cannot be distinguished by the eavesdropper (Eve). However, Alice and Bob know not only the noise levels in the channel but also the resistance value of their own resistor connected to the line. So, they can logically deduce the state at the other side from the resultant of the resistance values (deduced from channel noise intensity) and from state of their own switch.

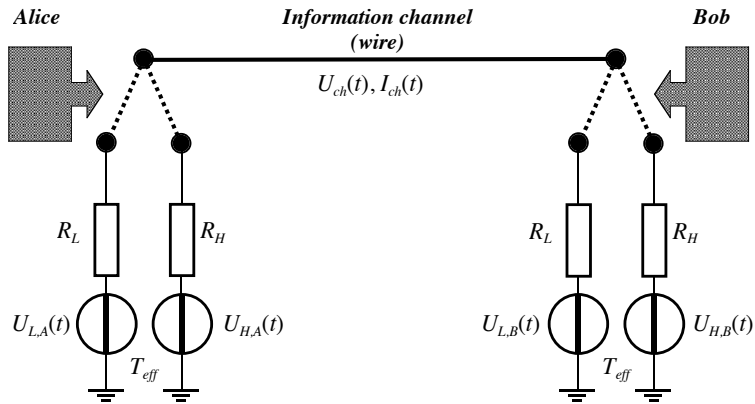

**Figure 1.** Basic circuitry of the KLJN secure key exchange system [9,10]. This simple version is not protected against active (invasive) attacks or attacks that utilize non-ideal features of the buildig elements.

This system, when ideal, offers unconditional (information theoretic) security against passive (listening) attacks only. To achieve unconditional security against active (invasive) attacks and/or attacks that utilize non-ideal features of the building elements, the following extra elements are essential [9,13,14]:
*a)* External noise generator to have stable noise temperature with high accuracy.

*b)* Filters to limit the frequency range to the measurement bandwidth and to the no-wave limit.

*c)* Measurement of the instantaneous voltage and current values at the two ends (by Alice and Bob) and to compare them via an authenticated public channel. When they differ, those situations may indicate invasive eavesdropping and those bits must be discarded. For the number of secure bits used up by the authentication, see Section 3.

While there were several attacks published against the KLJN system by utilizing non-ideal building elements; the above described defense method has been able to protect the system and maintain the unconditional security at the practically-perfect level [13]. Recently seven new KLJN schemes have been proposed which achieve the practically-perfect unconditional security with higher bit exchange rate.

The KLJN system has been proposed to provide unconditional security for hardware in computers, games and instruments [15].

*2.2 Noise-based logic; its hyperspace; and the string verification method*

Noise-based logic [16-26] is a new class of deterministic multivalued logic schemes where the information carrier is a system of random orthogonal time functions, as discussed below. For example, the stochastic time functions can have continuum amplitude distribution; they can be random spikes; or random telegraph signals. The universality of the binary versions of these logics schemes is proven.

Here we focus on the random telegraph wave scheme [22,23,26] and briefly introduce it based on [26]. There is a set of a simple type of random telegraph waves (RTW), $R_i(t_j)$, with discrete amplitudes $\pm 1$ and discrete time, where the index *i* stands for the *i*-th RTW and the index *j* for the *j*-th clock cycle. At the beginning of each clock period, a given random telegraph wave takes 1 or -1 amplitude with 50% probability. There is no memory in the system, except that the chosen amplitude is held until the end of the clock period where a new random selection takes place. Thus, for the sake of simplicity, we drop the continuum time parameter in the notation for an RTW, and refer only to the amplitude of the *i*th RTW during the *j*th clock cycle by $R_i(j)$. In other word,and RTW can be generated by a discrete random number generator that generates +1 or -1 with 50% probability each.

The product of an arbitrary number of independently generated RTWs :

$$W_x(j) = \prod_{i=1}^{N} R_i(j) \tag{1}$$

is called a hyperspace vector, which is also an RTW with the same statistical properties, and it is orthogonal to the original RTWs or any other RTW generated independently, i.e.:

$$\langle W_x(j) R_k(j) \rangle = \left\langle R_k(j) \prod_{i=1}^{N} R_i(j) \right\rangle = 0 \ , \tag{2}$$

where either $k \in \{1,...,N\}$ or $k > N$, which means that $R_k$ is an RTW generated independently of $R_1, R_2, ..., R_N$.

The string verification method utilizes the fact that, if two RTWs are generated independently, the probability $P(m)$ that their amplitudes are identical over *m* clock steps is

$$P(m) = 2^{-m} \tag{3}$$

To represent an *N* long bit string by such a product (hyperspace vector) we need 2*N* independently generated RTW signals to assign a different RTW to the *H* and *L* values of the *N* bits.

Suppose, both Alice and Bob have an *N*-bit long string and they would like to detect if one or more of their bit values are different [26]. Then they use the same publicly agreed set of 2*N* independent random number generators and assign each of them to the publicly agreed relevant bit and its value. Then using the *N* RTWs (random number generators) that represent the bit values in the string, they make the RTW product described by Eq. 1. Let as call Alice's product $W_A(t)$ and Bob's one $W_B(t)$.

Suppose that the bit strings of Alice and Bob differ by at least a single bit value. Then at least one RTW in these products (see Eq. 1) is different.

Obviously, if the $W_A(t)$ and $W_B(t)$ products contain the same RTW elements, which means that the two strings are identical, the following relations hold:

$$W_A(t_j) - W_B(t_j) = 0 \qquad (4)$$

$$W_A(t_j) W_B(t_j) = 1 \qquad (5)$$

Thus, a single comparator device checking for non-zero values in the first case; or a multiplier and a comparator checking for negative values in the second case can verify if the assumption about identical strings have been violated [26].

In conclusion, if the strings are different, the generated hyperspace vectors will also be different or deviate in a short time. Thus the comparison of short sample of $W_A(t)$ and $W_B(t)$ is enough to detect any difference with high probability. The probability that the difference of the strings is not detected is given by Eq. 3. This error probability gets infinitesimally small very quickly. For example, to reach the theoretical error probability of appr. $10^{-25}$ of the logic gates in computer chips at idealistic conditions, the required cock steps of comparison $m$ is only 83 (because $0.5^{83} \approx 10^{-25}$). Thus, Alice and Bob are able to detect that there is an arbitrary difference between their bit strings *of arbitrary length* of $N$ by communicating only 83 bits [26].

Note, an analogous representation of RTWs and the string verification can be achieved by using 0 and 1 amplitudes in the RTW and applying the XOR logic operation instead of multiplication.

After the above preparations, the description of the three different PUF systems is straightforward.

**3. Ultra-strong PUF hardware key with KLJN and one-time-pad**

The system is illustrated in Fig. 2. Both the Key and the Lock are integrated on chips. The manufacturer prepares the PUF with a secure key, which is cloneable. When the user uses the key first time (initialization) the Lock recognize the key by this cloneable secure key and then the Lock and the Key performs a KLJN key exchange and generates a new key that is unknown for even the manufacturer. The new secure key is save in a flash memory in the chip thus it is virtually impossible to access this information.

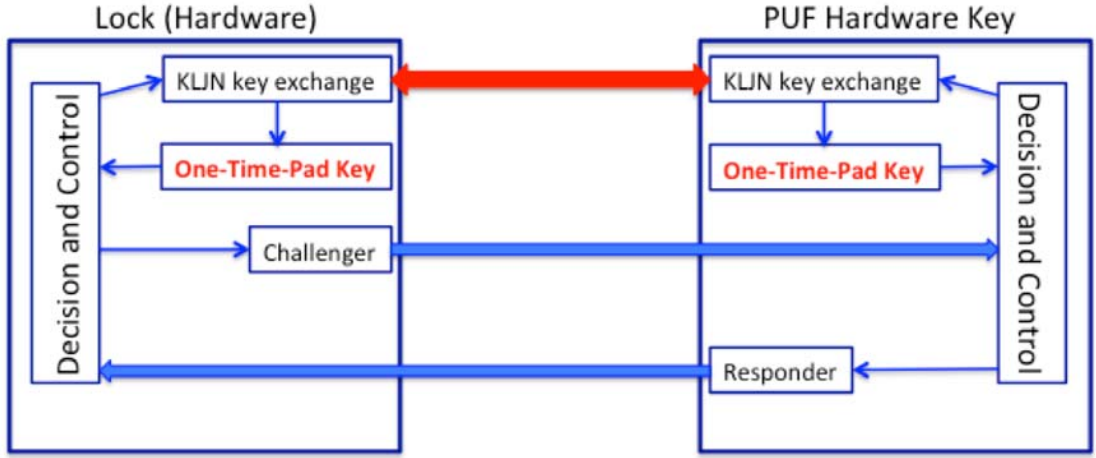

**Figure 2**. The Ultra-strong PUF hardware key with unconditionally secure KLJN key exchange and one-time pad.

After this initialization procedure, the uncloneable key is ready to be used and it is *ultra-strong* because at each new application of the key, they Lock and the Key again exchange a new independent unconditionally secure key. This PUF is ultra-strong because, if a counterfeiter accidentally succeeds to make a clone (this has exponentially low probability), that clone will be useless after next time the legal user applies the PUF key.

The challenge/response process is extremely simple: the Lock requests the PUF key to send the secure key they exchanged during last use of the key. It is important to note that the exchanged secure key is used as a *one-time pad*, which is the most secure way of secure communications, because during the next application of the PUF, already a new independent secure key will be used.

If a counterfeiter is trying to open the lock, his chance to succeed with a randomly generated *N*-bit long key is $2^{-N}$, where *N* is the length of the secure key, that is exponentially small. It is also important to not that there is no information about the next key and no modeling/simulation of system behavior can provide any information.

Note: minimum $N + \log_2 F$ key bits must be exchanged, where *F* is the number of authenticated public bits exchanged for current-voltage comparison against active/invasive attacks, because the authentication of *F* bits in a public channel requires using at least $\log_2 F$ secure bits [27]. The logarithmic dependence guarantees that the load by authentication stays negligibly small.

**4. Strong PUF hardware key with KLJN and noise-based logic**

Here we present a simplified version of the PUF key described above, which is still "strong", that is, even the manufacturer is unable to clone it after its first use, see Fig 3. The key refreshment is used only first time, during the initialization (which can be repeated if needed). Then during the subsequent applications the secure key cannot be used as a one-time-pad because the same secure key must be used. To provide and exponentially difficult job for the counterfeiter the challenge must proceed via a secure communication where the secure key is used in the cipher. The Lock generates a random string, which then gets encrypted by the cypher and sent to the PUF key. The response system in the PUF key utilizes noise-based logic (NBL) and it generates and sends the product RTW described by Eq. 1, that corresponds to the given bit string. A counterfeiter has again $2^{-N}$ chance to succeed if he tries with his only available tool: brute force.

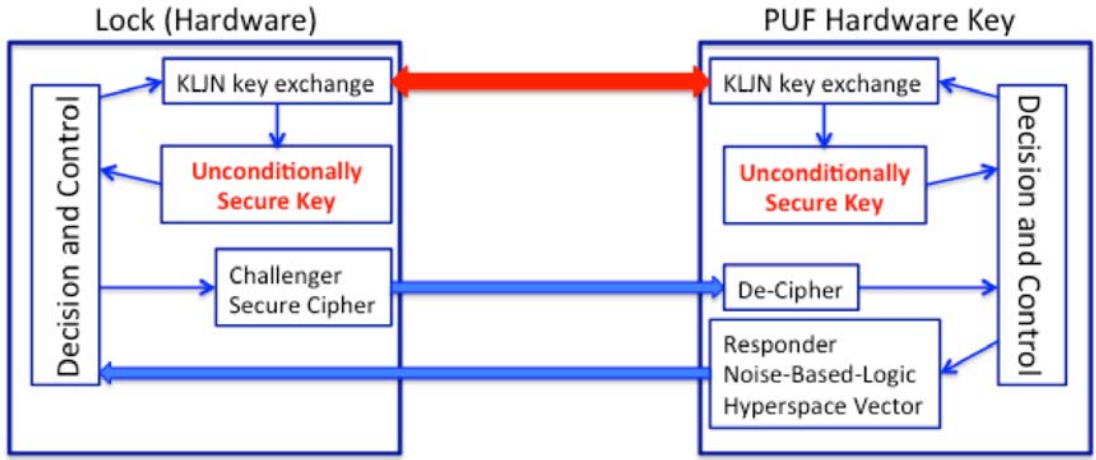

**Figure 3**. The Strong PUF hardware key with unconditionally secure KLJN key exchange and noise-based logic.

Then the Lock will use the string verification method described above: it also generate the required RTW product and uses Eqs. 4 and 5 to check if the sequence coming from the Key is satisfactory. If the Key sends $N$ clock steps of the product, the chance that a brute force counterfeiter escapes detection is again $2^{-N}$.

The outline of the Responder is shown in Figure 4. The secure key bits shared by the unconditionally secure KLJN protocol determine for each bit if its pseudo-random-number generator A or B is assigned to its high or low value. All the $2N$ generators are independent and publicly known except their secret assignment.

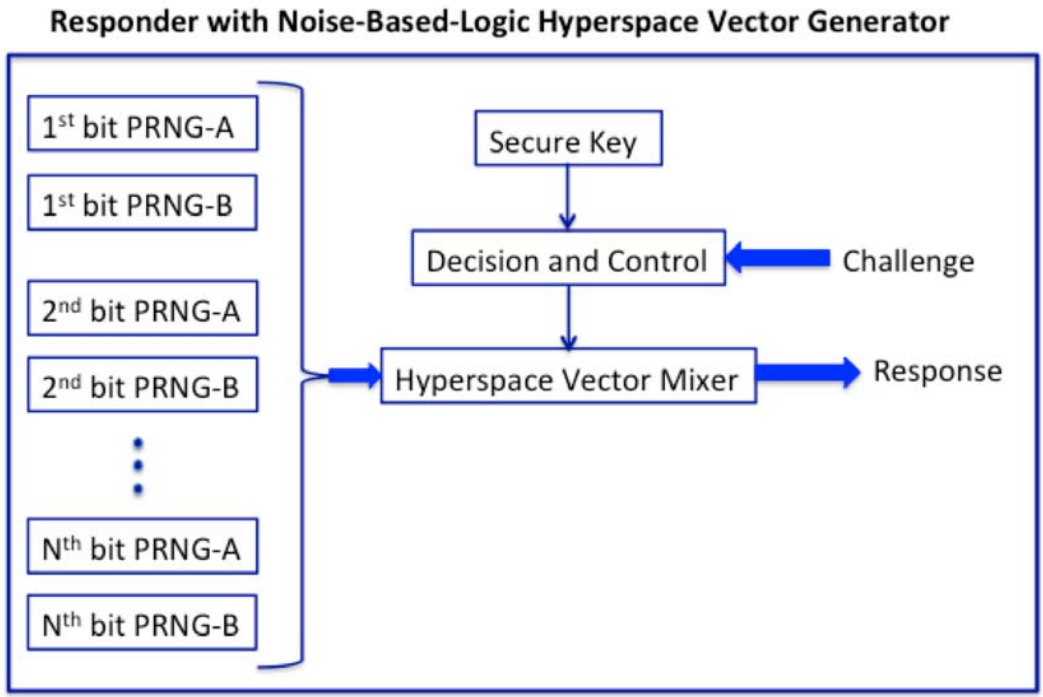

**Figure 4.** The Responder with RTW-based noise-based logic.

## 5. Simple PUF hardware key with noise-based logic

The simplest PUF utilizing noise-based informatics is shown in Fig. 5.. It has a secure key that is implemented by the manufacturer thus it can be cloned (this means it is not a strong version) however this PUF cannot be cloned by anybody else. Similarly to the former system, the challenge-response system utilizes secure communication and the NBL scheme shown in Fig 4. and described in Section 2.2.

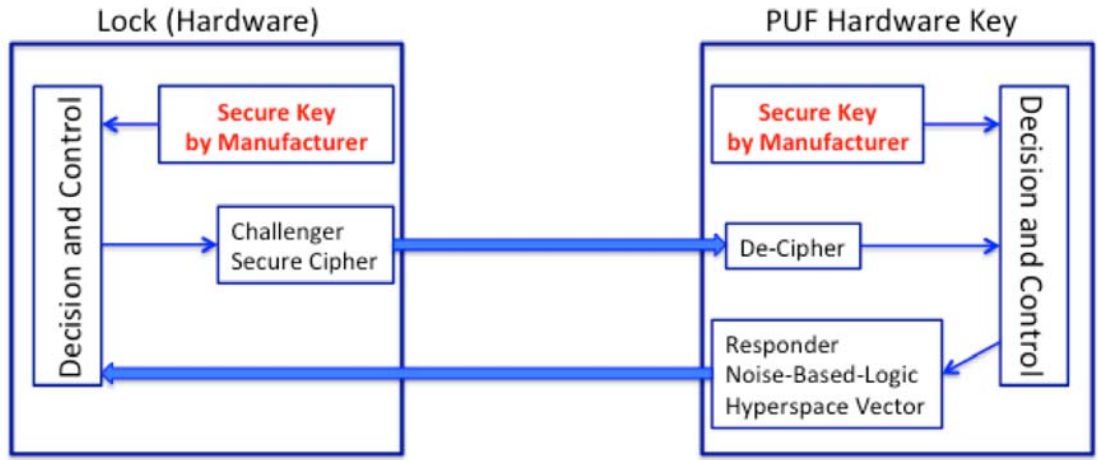

**Figure 5.** Simple PUF hardware key with noise-based logic

## 6. Conclusion

We showed three ways to realize physical uncloneable function (PUF) encryption keys where noise-based informatics is utilized. The key lengths that can easily be much longer than 256 bits. All these systems can be integrated on a chip, which provides robustness against tampering and environmental effects.